\documentclass[11pt, a4paper, onecolumn]{shefaire}

\usepackage[a4paper]{geometry}
\usepackage[numbers, sort&compress, square]{natbib}
\usepackage{xspace}  %
\usepackage{siunitx}  %
\usepackage{amsthm, amsmath, amsfonts, amssymb}
\usepackage{placeins}
\usepackage[english]{babel}
\usepackage{cmap}
\usepackage[T1]{fontenc}
\usepackage{url}
\usepackage{titletoc}
\usepackage[title]{appendix}

\usepackage[colorlinks=true, allcolors=blue]{hyperref}
\usepackage[capitalise,noabbrev]{cleveref}
\usepackage{pifont}
\usepackage[dvipsnames]{xcolor}
\usepackage{textcomp} %
\usepackage[overload]{textcase}
\usepackage{graphicx}
\usepackage{colortbl}
\usepackage{booktabs}
\usepackage{changepage}
\usepackage{enumitem} %
\usepackage{tabularx}
\usepackage{datetime}
\usepackage{fancyhdr}  %
\usepackage{lastpage}  %
\usepackage{enumitem}
\usepackage[explicit]{titlesec}
\usepackage{bibentry}
\usepackage{mdframed}
\usepackage{caption}
\usepackage{needspace}
\usepackage{siunitx}
\usepackage{lscape}
\usepackage{float}
\usepackage{lipsum}
\usepackage{pdfpages}
\usepackage{xcolor}
\usepackage[figuresright]{rotating}
\usepackage[none]{hyphenat}
\usepackage{hyperref}
\usepackage{bbding}
\usepackage{datetime2}
\usepackage{array}
\usepackage{multirow}
\usepackage{MnSymbol}
\newcolumntype{C}[1]{>{\centering\arraybackslash}p{#1}}
\newcommand{\newcheck}{$\checkmark$}
\newcommand{\revcheck}{$\color{blue}\checkmark$}

\newcommand{\rev}[1]{{\leavevmode\color{black}#1}}
\newcommand{\add}[1]{{\color{black}#1}}

\newlength{\wideitemsep}
\let\olditem\item
\renewcommand{\item}{
    \setlength{\parskip}{2ex}
    \setlength{\itemsep}{\wideitemsep}
    \olditem
}

\title{Interpretable Multimodal Learning for Tumor Protein-Metal Binding: Progress, Challenges, and Perspectives} 

\correspondingauthor{sangwei@sxmu.edu.cn, h.lu@sheffield.ac.uk}

\author[1,2,3]{Xiaokun Liu}
\author[4]{Sayedmohammadreza Rastegari}
\author[5]{Yijun Huang}
\author[6]{Sxe Chang Cheong}
\author[1,2]{Weikang Liu}
\author[1,2]{Wenjie Zhao}
\author[1,2]{Qihao Tian}
\author[1,2]{Hongming Wang}
\author[1,2,3]{Yingjie Guo}
\author[7,8]{Shuo Zhou}
\author[7]{Sina Tabakhi}
\author[7,8]{Xianyuan Liu}
\author[1,3]{Zheqing Zhu}
\author[9,10,*]{Wei Sang}
\author[1,7,8,*]{Haiping Lu}

\affil[1]{Institute of Big Data Science and Industry, Shanxi University, Taiyuan, China}
\affil[2]{School of Computer and Information Technology, Shanxi University, Taiyuan, China}
\affil[3]{Key Laboratory of Evolutionary Science Intelligence of Shanxi Province, Taiyuan, Shanxi, China}
\affil[4]{Faculty of Computer Engineering, University of Isfahan, Isfahan, Iran}
\affil[5]{The First Clinical Medical School, Shanxi Medical University, Taiyuan, China}
\affil[6]{School of Medicine \& Population Health, University of Sheffield, Sheffield, UK}
\affil[7]{School of Computer Science, University of Sheffield, Sheffield, UK}
\affil[8]{Centre for Machine Intelligence, University of Sheffield, Sheffield, UK}
\affil[9]{Department of Biochemistry and Molecular Biology, School of Basic Medical Sciences, Shanxi Medical University, Taiyuan, China}
\affil[10]{Institute of Medical Technology, Shanxi Medical University, Taiyuan, China}

\begin{abstract}

In cancer therapeutics, protein-metal binding mechanisms critically govern the pharmacokinetics and targeting efficacy of drugs, thereby fundamentally shaping the rational design of anticancer metallodrugs. While conventional laboratory methods used to study such mechanisms are often costly, low throughput, and limited in capturing dynamic biological processes, machine learning (ML) has emerged as a promising alternative. Despite increasing efforts to develop protein-metal binding datasets and ML algorithms, the application of ML in tumor protein-metal binding remains limited. Key challenges include a shortage of high-quality, tumor-specific datasets, insufficient consideration of multiple data modalities, and the complexity of interpreting results due to the ``black box'' nature of complex ML models. This paper summarizes recent progress and ongoing challenges in using ML to predict tumor protein-metal binding, focusing on data, modeling, and interpretability. We present multimodal protein-metal binding datasets and outline strategies for acquiring, curating, and preprocessing them for training ML models. Moreover, we explore the complementary value provided by different data modalities and examine methods for their integration. We also review approaches for improving model interpretability to support more trustworthy decisions in cancer research. Finally, we offer our perspective on research opportunities and propose strategies to address the scarcity of tumor protein data and the limited number of predictive models for tumor protein-metal binding. We also highlight two promising directions for effective metal-based drug design: integrating protein-protein interaction data to provide structural insights into metal-binding events and predicting structural changes in tumor proteins after metal binding.

\end{abstract}

\begin{document}
\maketitle

\renewcommand{\thesection}{\arabic{section}}
\renewcommand{\thefigure}{\arabic{figure}}
\renewcommand{\thetable}{\arabic{table}}
\renewcommand{\theequation}{\arabic{equation}}

\section*{Highlights}
\begin{itemize}[leftmargin=*]

\item Tumor protein-metal binding is key to designing metal-based anticancer therapies.
\item Multimodal data can enable accurate binding prediction and deep biological insights.
\item Cross-referencing datasets can help address the scarcity of tumor protein-metal-binding data.
\item Existing protein–metal binding models can be adapted for tumor-specific prediction.
\item Improving interpretability can support cross-disciplinary research and collaboration.

\end{itemize}

\section*{Keywords}
\textbf{Tumor Protein-Metal Binding; Multimodal Learning; Interpretability; Data Integration}

\section{Introduction}

Cancer remains one of the leading causes of death worldwide, accounting for approximately 25\% of all fatalities. Each year, about 0.5\% of the global population is newly diagnosed with the disease \cite{mustapha2022cancer}. In 2022 alone, nearly 20 million new cases were reported, resulting in an estimated 9.7 million cancer-related deaths \cite{bray2024global}. While chemotherapy has long been a primary treatment option, its use is often limited by severe side effects, such as nausea and systemic complications \cite{mustapha2022cancer}. 
Therefore, there is a need to improve chemotherapy by reducing toxicity and enhancing efficacy. Metal-based drugs, owing to their diverse chemical compositions, have emerged as a promising class of chemotherapeutic agents.
By changing the metal center, ligands, and metal-ligand interactions, researchers can optimize metal-based compounds to match the biological characteristics of different cancer cells. This adaptability supports the development of more personalized treatment strategies. Widely used examples, such as cisplatin and carboplatin, have demonstrated success in overcoming chemoresistance \cite{pena2022metallodrugs}. Metal-based drugs can also be integrated with nanotechnology and immunotherapy to improve their therapeutic effectiveness. For example, metal-intermetallic compounds have shown potential not only in eliminating cancer cells but also in stimulating memory immune cells, thereby reducing the risk of tumor relapse \cite{zhu2024intermetallics}.

\begin{figure}[!ht] %
    \centering
    \includegraphics[width=\textwidth]{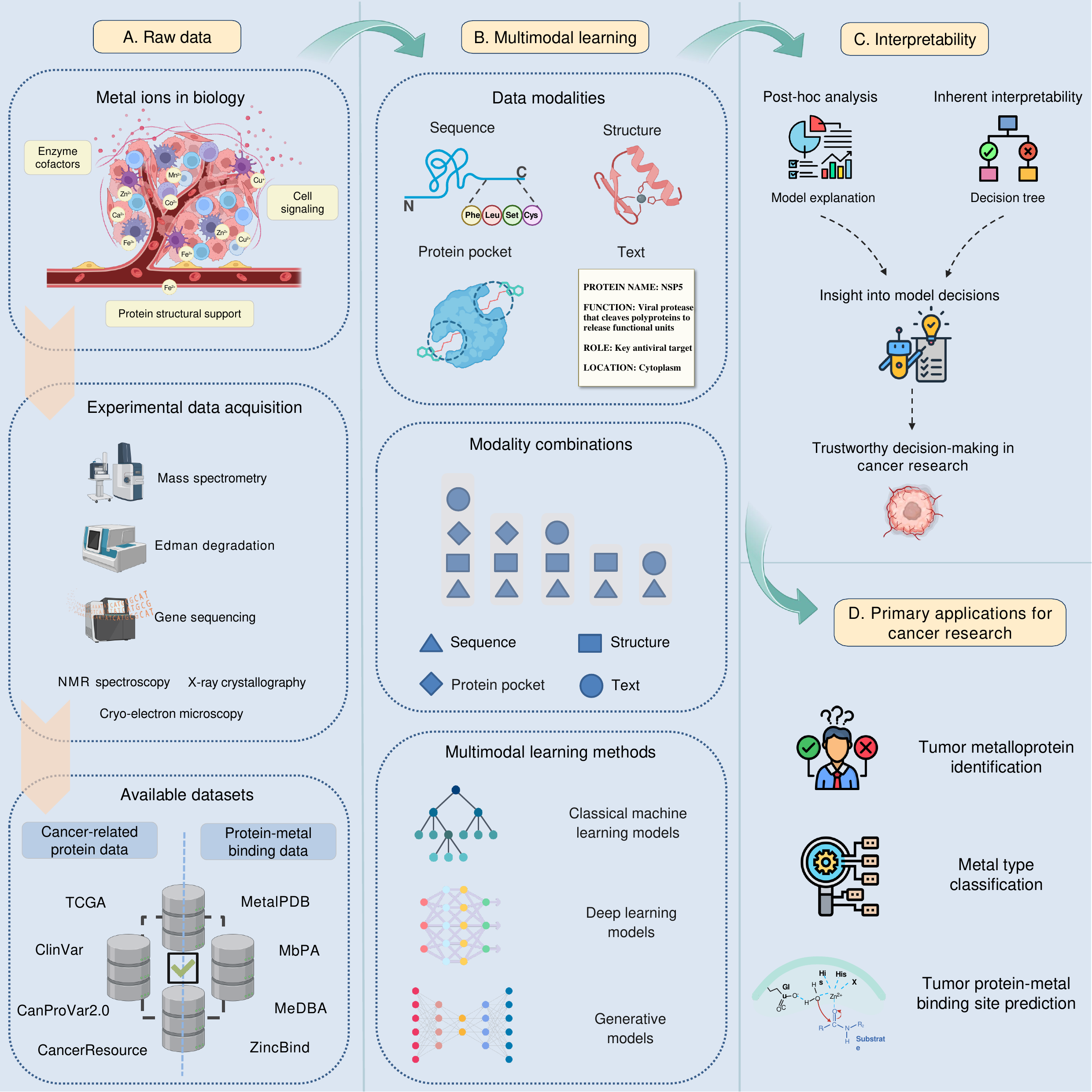} %
    \caption{Interpretable multimodal learning for tumor protein-metal binding. A) The process of transforming bioinformatics data to ML-oriented datasets. B) The data modalities, their combinations, and the commonly-used multimodal learning methods. C) The key roles of model interpretability. D) The primary applications for cancer research. Created with \href{https://BioRender.com}{BioRender}.} %
    \label{fig:Fig.1} %
\end{figure}

The design and characterization of metal-based compounds mainly rely on conventional wet-lab methods, such as mass spectrometry, X-ray crystallography, and nuclear magnetic resonance \cite{hartinger2013application,song2024probing,messori2017protein,jensen2007investigating}, as shown in Fig. \ref{fig:Fig.1}A. These methods provide valuable insights into the properties of metal-based compounds and the structural basis of their interactions with tumor proteins. However, they are often expensive, time-consuming, and low throughput, and they struggle to capture dynamic biological processes. 
The emergence of machine learning (ML) presents new opportunities, as ML techniques can efficiently analyze large amounts of complex biological data and build predictive models for protein-metal binding, thereby accelerating the discovery of metal-based drugs. Figure \ref{fig:Fig.1}D shows three key ML application tasks in tumor protein-metal binding: tumor metalloprotein identification, metal type classification, and binding site prediction. Despite this potential, the application of ML remains limited due to challenges including the scarcity of high-quality datasets, insufficient integration of multimodal information, and a lack of interpretability in existing models.

To summarize recent advances, highlight key challenges, and explore promising directions, this paper is organized around three key areas: data resources, multimodal learning approaches, and interpretability techniques. First, we review existing data resources and observe that, while high-quality general protein-metal binding datasets are abundant, those specific to tumor proteins are scarce. Since tumor proteins are not structurally different from general proteins, the learning mechanisms in ML models remain applicable across both types. Therefore, we collect datasets on both general protein-metal binding and tumor proteins. We then propose a strategy to construct tumor-specific datasets by integrating these complementary sources. For example, although the Protein Data Bank (PDB) \cite{burley2017protein} does not explicitly annotate tumor protein-metal binding interactions, it includes cancer-associated proteins with experimentally resolved metal-binding structures. Identifying and using these entries to extract key biological features can help mitigate current data scarcity. 

Next, different data modalities can provide complementary biological insights to enhance feature learning when effectively integrated. To capture this diversity, we identify four key data modalities relevant to tumor protein: sequence, structure, protein pocket, and text, as shown in Fig. \ref{fig:Fig.1}B. For each modality, we introduce its unique characteristics and present an ML-oriented data processing workflow, including data acquisition, data preprocessing, and feature extraction. This workflow can support researchers in locating, preparing, and adopting raw data for ML model development. 

Integrating these heterogeneous modalities into a model is nontrivial and can significantly influence model performance. To better understand the current landscape, we examine existing ML models developed for tumor protein-metal binding. To the best of our knowledge, only three such models have been developed to date - multichannel convolutional neural network (MCCNN) \cite{koohi2019predicting}, MetalPrognosis \cite{jia2024metalprognosis}, and MetalTrans \cite{zhang2024metaltrans} - but they all have limited success in multimodal integration. Therefore, we study multimodal ML methods from broader bioinformatics applications, including 14 recent protein–metal binding models and four protein foundation models, and propose strategies for adapting them to tumor-specific tasks. Building on these insights, we highlight the important value of multimodality for tumor protein-metal binding, suggest promising combinations of data modalities, and provide an overview of general multimodal learning methods, as shown in Fig. \ref{fig:Fig.1}B.

Finally, in medical and biological applications, predictive accuracy alone is insufficient. ML models should also be interpretable and trustworthy, supporting scientific and clinical understanding. However, the ``black box'' nature of popular complex ML models limits transparency into their internal decision-making processes. In practice, interpretability is often limited to indirect evaluations, such as performance comparisons with benchmark models, without clearly explaining how specific inputs contribute to predictions. This lack of transparency poses challenges for researchers, particularly those from clinical and biological backgrounds, in assessing model reliability and extracting clinically or biologically relevant insights. To address this issue, we examine interpretability in the context of tumor protein-metal binding from two perspectives: designing models with inherent interpretability and applying post hoc analysis techniques.

Building on these findings, we summarize key challenges and discuss perspectives on potential solutions. Moreover, we identify two emerging directions that can advance the prediction of tumor protein–metal binding. The first involves integrating protein-protein interaction data to provide a structural context for metal-binding events. The second focuses on predicting the structural changes in tumor proteins after metal binding, which can inform the design of more effective metal-based therapies. To our knowledge, this is the first comprehensive study of interpretable multimodal ML for tumor protein-metal binding. Our work synthesizes research advances, identifies ongoing challenges, and offers a forward-looking perspective into future research opportunities. By integrating knowledge from protein structure datasets, multimodal ML, and interpretability analysis, we aim to support the development of biologically grounded, accurate, and clinically applicable ML models for tumor protein-metal binding prediction.

The main contributions of this paper are threefold:

\begin{itemize}
    \item We present potential data resources for tumor protein-metal binding, a systematic data processing workflow, and a strategy to construct tumor-specific datasets by integrating protein-metal binding and cancer-related datasets.
    \item We identify key protein data modalities, explore their combinations, and review multimodal learning methods focusing on their applicability to tumor protein-metal binding. We also emphasize the importance of interpretability and summarize current approaches that support biological and clinical insight.
    \item We offer a forward-looking perspective by proposing two promising research directions, i.e., exploring protein-protein interaction and predicting tumor protein structural changes upon metal binding to inform drug design and discovery.
\end{itemize}

The remainder of this paper is organized as follows. Section \ref{sec:sec2} introduces the role of metal ions in tumor proteins, and Section \ref{sec:sec4} focuses on tumor protein-metal binding data sources and data processing workflow. Section \ref{sec:sec5} studies related multimodal learning resources and methods, and Section \ref{sec:sec6} investigates the importance of model interpretability and related methods for interoperability improvement. Section \ref{sec:sec7} summarizes key challenges and perspectives, highlighting two further directions for advancing tumor protein-metal binding research. Finally, Section \ref{sec:sec8} concludes this work.

\section{Biological significance of tumor protein-metal binding}
\label{sec:sec2}

The tumor microenvironment is a dynamic and essential foundation for tumor survival, composed of diverse cells, vasculature, secreted factors, and extracellular matrix components that tumors actively shape and induce. Driven by their intense metabolic demands, tumors generate unique conditions, such as hypoxia, acidity, and elevated reactive oxygen species (ROS) that impair the function of pro-inflammatory immune cells, resulting in evasion of the immune system \cite{chao2023biomaterials}. Within the tumor microenvironment, metal ions engage with proteins to establish a sophisticated and multi-tiered regulatory network that precisely modulates cellular signaling pathways, metabolic activities, and immune cell functionalities. This elaborate network plays a pivotal role in governing tumor initiation and progression. 

Calcium ions ($\text{Ca}^{2+}$) bind to the EF-hand domains of calmodulin \cite{crotti2013calmodulin}, inducing a conformational change that enables the $\text{Ca}^{2+}$-calmodulin complex to activate calcineurin. This, in turn, dephosphorylates the nuclear factor of activated T cells (NFAT), promoting its translocation to the nucleus and subsequent transcription of cytokine genes such as IL-2 \cite{gao2024regulation}. In tumor cells, $\text{Ca}^{2+}$ signaling regulates key pathways such as PI3K/AKT and MAPK, which are critical for cell survival and proliferation. Zinc ions ($\text{Zn}^{2+}$) interact with zinc finger proteins such as MTF-1, binding to conserved cysteine and histidine residues within their zinc finger domains. This interaction regulates the expression of metallothioneins (MT1/2), which protect tumor cells from oxidative stress. In T cells, $\text{Zn}^{2+}$ modulates the activity of transcription factors like GATA-3 and T-bet, thus influencing Th1/Th2 differentiation and antitumor immunity \cite{gao2024regulation}.

Iron ions ($\text{Fe}^{2+}/\text{Fe}^{3+}$) bind to iron-sulfur clusters in proteins such as aconitase \cite{martins2018persistence} and SDHB \cite{plugge2012complete}, regulating mitochondrial metabolism and ROS production. Dysregulated iron metabolism can induce ferroptosis in tumor cells through the accumulation of lipid peroxides \cite{tsuji20231}. In TAMs, $\text{Fe}^{2+}/\text{Fe}^{3+}$ modulates polarization by regulating HIF-1$\alpha$ and NF-$\kappa$B signaling, thereby influencing tumor progression \cite{barman2016fe}. Copper ions ($\text{Cu}^{+}/\text{Cu}^{2+}$) bind to copper-dependent enzymes such as superoxide dismutase 1 (SOD1) and LOX, modulating intracellular redox reactions and extracellular matrix remodeling. Manganese ions ($\text{Mn}^{2+}$) activate manganese superoxide dismutase (Mn-SOD) by binding to its active site, which is composed of conserved histidine and aspartate residues. This binding induces a conformational change that enables Mn-SOD to catalyze the dismutation of superoxide radicals ($\text{O}_{2}^{-}$) into hydrogen peroxide ($\text{H}_{2}\text{O}_{2}$) and oxygen ($\text{O}_{2}$) \cite{borgstahl1992structure,fridovich1981superoxide}, thereby reducing oxidative stress and protecting tumor cells from apoptosis. Magnesium ions ($\text{Mg}^{2+}$) bind to phosphate groups of ATP, forming a Mg-ATP complex that stabilizes ATP and enhances its biological availability for energy transfer and hydrolysis. In tumor cells, $\text{Mg}^{2+}$ acts as a cofactor for kinases such as mitogen-activated protein kinase (MAPK), including ERK, p38, and JNK, facilitating their phosphorylation and activation. This regulation of MAPK signaling pathways influences critical cellular processes, including proliferation, differentiation, and apoptosis.

These protein-metal interactions form a highly coordinated regulatory system that influences tumor progression, immune response, and cellular metabolism. The biological significance underscores the potential for developing targeted cancer therapies, motivating this work.

\begin{table}[t]
\centering
\caption{Cancer-related protein datasets}
\resizebox{\textwidth}{!}{%
\begin{tabular}{l l l}
\toprule \toprule
\textbf{Dataset} & \textbf{Key feature} & \textbf{URL} \\
\midrule
TCGA \cite{tomczak2015review}    & Cancer genomics & \url{cancer.gov/ccg/research/genome-sequencing/tcga} \\
CanProVar2.0 \cite{zhang2017canprovar}     & Cancer protein variants & \url{canprovar2.zhang-lab.org/} \\
ClinVar \cite{landrum2016clinvar} & Clinical variant data (includes cancer-related variants) & \url{ncbi.nlm.nih.gov/clinvar/} \\
CancerResource \cite{gohlke2016cancerresource} & Cancer-relevant proteins, mutations and drug interactions & \url{bioinformatics.charite.de/cancerresource/} \\
UALCAN \cite{chandrashekar2022ualcan}  & Cancer omics analysis & \url{ualcan.path.uab.edu/} \\
CPTAC \cite{edwards2015cptac}   & Tumor proteomics & \url{proteomics.cancer.gov/programs/cptac} \\
HuVarBase \cite{ganesan2019huvarbase}     & Human variants at protein and gene levels & \url{iitm.ac.in/bioinfo/huvarbase} \\
\bottomrule \bottomrule
\end{tabular}
}
\label{tab:CancerDatabases}
\end{table}

\begin{table}[t]
\centering
\caption{Protein-metal binding datasets}
\resizebox{\textwidth}{!}{%
\begin{tabular}{l l l c c c c}
\toprule \toprule
\multirow{2}{*}{\textbf{Dataset}} & \multirow{2}{*}{\textbf{Key feature}} & \multirow{2}{*}{\textbf{URL}} & \multicolumn{4}{c}{\textbf{Data modality}} \\
\cmidrule(lr){4-7}
 &  &  & \textbf{Sequence} & \textbf{Structure} & \textbf{Pocket} & \textbf{Text}\\
\midrule
UniProt \cite{uniprot2025uniprot} & Protein sequences and functional information & \url{uniprot.org} & \(\checkmark\) & & & \(\checkmark\)  \\
PDB \cite{burley2017protein} & Protein 3D structures with metal coordination  & \url{rcsb.org} & \(\checkmark\) & \(\checkmark\) & \(\checkmark\) & \(\checkmark\) \\
MetalPDB \cite{putignano2018metalpdb} & Metal sites in proteins & \url{metalpdb.cerm.unifi.it} & \(\checkmark\) & \(\checkmark\) & \(\checkmark\) & \(\checkmark\) \\
MbPA \cite{li2023metal} & Metalloprotein data of multiple species & \url{bioinfor.imu.edu.cn/mbpa} & \(\checkmark\) & \(\checkmark\) & \(\checkmark\) & \(\checkmark\) \\
BioLip2 \cite{zhang2024biolip2} & Ligand-protein interactions including metals & \url{zhanggroup.org/BioLiP} & \(\checkmark\) & \(\checkmark\) & \(\checkmark\) & \(\checkmark\) \\
Q-BioLip \cite{10.1093/gpbjnl/qzae001} & Quaternary structure-based protein-ligand interactions & \url{yanglab.qd.sdu.edu.cn/Q-BioLiP} & \(\checkmark\) & \(\checkmark\) & \(\checkmark\) & \(\checkmark\) \\
MESPEUS \cite{lin2024mespeus} & Experimental metal-binding data from crystallography & \url{mespeus.nchu.edu.tw} &  & \(\checkmark\) & \(\checkmark\) & \(\checkmark\) \\
InterMetalDB \cite{tran2021intermetaldb} & Metal ion binding and functional roles & \url{intermetaldb.biotech.uni.wroc.pl} &  & \(\checkmark\) & \(\checkmark\) & \(\checkmark\) \\
ZincBind \cite{ireland2019zincbind} & Zinc-binding sites & \url{zincbind.net} &  & \(\checkmark\) & \(\checkmark\) & \(\checkmark\) \\
PocketDB \cite{bhagavat2018augmented} & Protein pockets & \url{proline.biochem.iisc.ernet.in/PocketDB/} &  &  & \(\checkmark\) & \(\checkmark\) \\
CavitySpace \cite{wang2022cavityspace} & Protein cavities and pockets identified from 3D structures & \url{pkumdl.cn:8000/cavityspace} &  &  & \(\checkmark\) & \(\checkmark\) \\

\bottomrule \bottomrule
\end{tabular}
}
\label{tab:protein-metal binding datasets}
\end{table}

\section{Tumor protein-metal binding data}
\label{sec:sec4}

Given the limited availability of tumor-specific protein-metal binding datasets, applying ML in this domain remains challenging. This section reviewed existing data resources, identified key data modalities, and proposed strategies to mitigate data scarcity. We introduced publicly available cancer-related and protein-metal binding datasets and discussed how integrating these resources can support the construction of tumor-specific datasets. Additionally, we examine four key data modalities in protein research and present a data processing workflow to prepare these modalities for ML modeling.

\subsection{Tumor protein and metal-binding datasets}
\label{subsec:data1}

Several publicly available datasets offer valuable information on cancer biology and protein-metal binding, enabling researchers to investigate the structural and functional roles of metal ions in both tumor and general proteins. As summarized in Table~\ref{tab:CancerDatabases} and \ref{tab:protein-metal binding datasets}, cancer-related datasets mainly focus on genomic, proteomic, and clinical data relevant to tumor biology, while protein-metal binding datasets contain detailed structural and functional annotations for proteins that coordinate metal ions. 
Because tumor proteins are not structurally distinct from general proteins, we propose a strategy to integrate tumor protein annotations with metal-binding data, enabling researchers to cross-reference information across specialized datasets.
By leveraging multiple data sources, researchers can identify proteins of interest and explore their metal coordination properties and relevance in the context of tumor biology.
For example, a protein can be identified through The Cancer Genome Atlas Program (TCGA) \cite{tomczak2015review}, and its metal-binding features, such as coordination details, binding motifs, and potential roles, can be explored using UniProt \cite{uniprot2025uniprot}. 
A case in point is the copper transport protein ATOX1, which exhibits altered RNA transcript levels in various cancers \cite{fu2013metallothionein}. Researchers can find ATOX1 in TCGA and use its UniProt accession number (O00244) to access its metal-binding information. ATOX1 binds to copper and functions as a chaperone, delivering copper to targets such as P-type ATPases. The dataset also links to studies on ATOX1’s roles in breast and liver cancer, while structural data is accessible via external resources, such as the PDB.

This strategy highlights how integrating datasets such as TCGA and PDB enables researchers to investigate metal-binding proteins, even when explicit metal annotations are missing in tumor datasets, by leveraging cross-referenced structural and functional insights.
Looking ahead, we envision a future where an integrated dataset combines gene expression, metal binding, and structural data, streamlining protein-level analysis in cancer. Such a resource would allow researchers to readily query proteins like ATOX1 to assess tumor-specific expression, metal coordination, and structural context, ultimately supporting the discovery of new targets for metal-based cancer therapies.

\subsection{Raw data acquisition}
\label{subsec:dtf}

\begin{figure*}[t]
    \centering
    \includegraphics[width=\columnwidth]{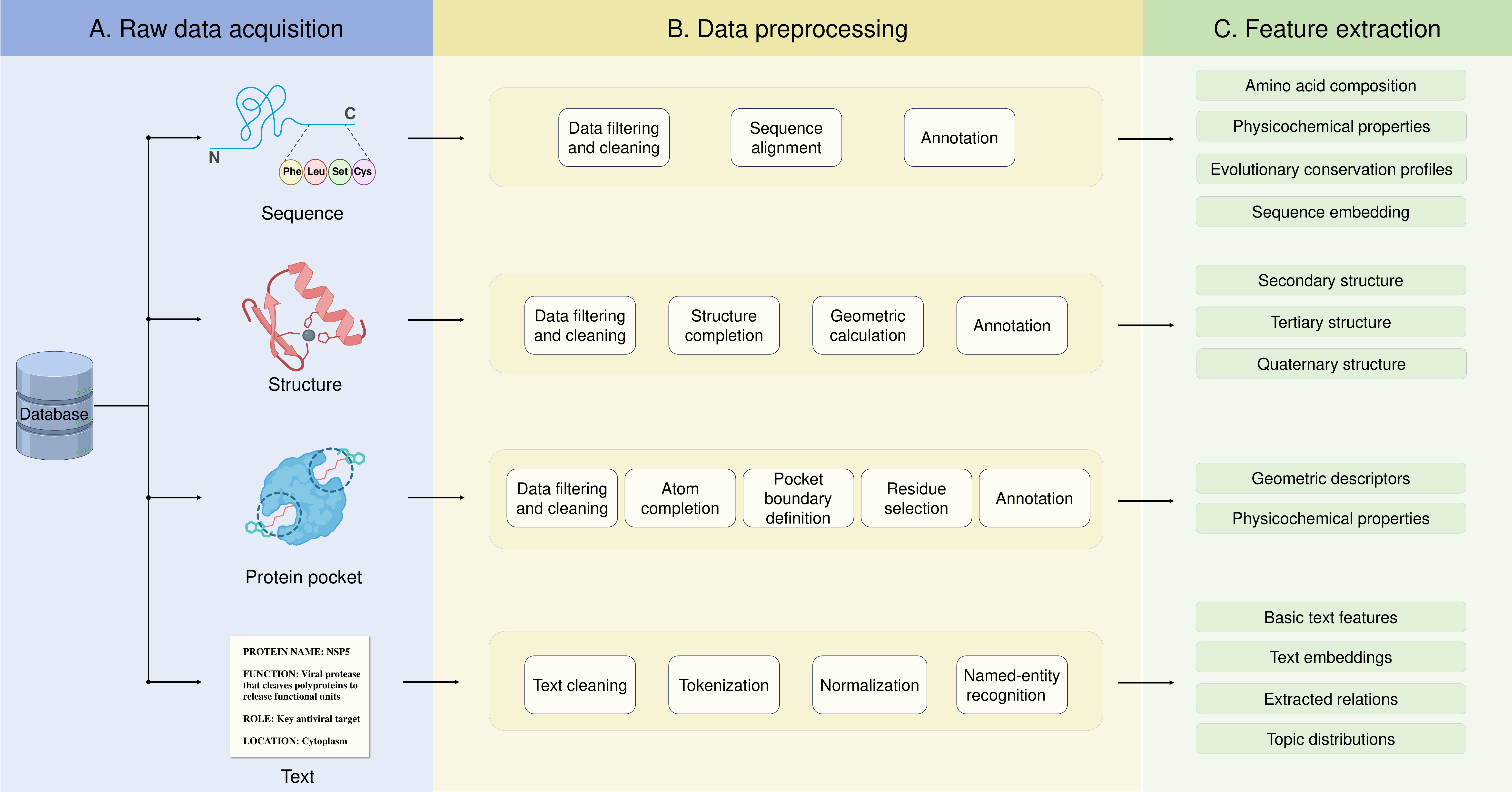}
        \caption{Overview of data processing workflow for four data modalities: sequence, structure, protein pocket, and text. This workflow progresses from raw data acquisition and data preprocessing to feature extraction. Created with \href{https://BioRender.com}{BioRender}.}
    \label{fig:ftd_fig}
\vspace{-5pt}
\end{figure*}

Since different data modalities provide complementary biological insights, integrating them can significantly enhance feature learning and understanding. In protein research, we identify four key data modalities (Fig. \ref{fig:ftd_fig}A): sequence, structure, protein pocket, and text. Each modality has distinct characteristics and unique biological information. Below, we introduce the key features of each modality and describe the methods used to prepare them for ML applications.

\textbf{Sequence} represents the unique order of amino acids in a protein chain, defining its primary structure. It reveals the specific arrangement of residues that can indicate functional regions, including conserved motifs-such as zinc fingers, H-N-H, and CXXC — that frequently contain key metal-binding residues like cysteine, histidine, aspartic acid, and glutamic acid, which coordinate metal ions critical for structural stability and function \cite{berg1990zinc, cheng2023co}. Standard file formats include FASTA, GenBank, and other plain text formats that list amino acids or nucleotide sequences. These formats can be derived from multiple datasets, such as the PDB and UniProt, through their websites or application programming interfaces (APIs). It is also possible to extract protein sequences from a PDB structural data file using tools like Biopython \cite{cock2009biopython}.

\textbf{Structure} provides a 3D view of a protein and related metal ions, detailing how its atoms are spatially organized into secondary, tertiary, or quaternary structures, and illustrates the folding pattern and conformational details critical for understanding protein-ligand interactions, including those with metal ions. Secondary structures comprise local folded regions formed by interactions among backbone atoms, so secondary structures of a protein refer to local, repeated arrangements of the amino acid chain, such as alpha-helices and beta-sheets, which are stabilized mainly by hydrogen bonds between the backbone atoms. Tertiary structures represent the overall 3D shape of a single polypeptide chain arising from interactions between the side chains (R groups) of the amino acids, including hydrogen bonds, ionic bonds, hydrophobic interactions, and disulfide bridges, which determine the protein's overall functionality. Quaternary structures apply to proteins composed of multiple polypeptide chains, describing the arrangement and interaction of these subunits. Structural data is typically obtained using X-ray crystallography, NMR spectroscopy, and cryo-electron microscopy (cryo-EM). This information is stored in formats like PDB files or mmCIF, which include 3D coordinates and other structural details. In metal-binding prediction, secondary and tertiary structures are most commonly utilized. Q-BioLip \cite{10.1093/gpbjnl/qzae001} is a recent dataset that provides quaternary structure-based protein-ligand interaction information.

\textbf{Protein pocket} is a 3D cavity within the complex architecture of protein structures that serves as a crucial microenvironment for molecular interactions. These indentations can accommodate various ligands, including metal ions, thereby mediating a vast array of protein functions. While pocket data can be extracted from protein structure datasets, several specialized datasets provide detailed pocket data, including PocketDB \cite{bhagavat2018augmented}, MetalPDB \cite{putignano2018metalpdb}, MetalMine \cite{nakamura2009metalmine}, and CavitySpace \cite{wang2022cavityspace}.

\textbf{Text} encompasses the detailed, human-readable information associated with biological entities, such as proteins, found in datasets and scientific literature. Unlike sequences or structures, text provides rich contextual insights, including functional annotations, experimental findings, and disease associations, which complement other data modalities \cite{verspoor2012text}. This modality captures the precise language used to describe protein characteristics, such as metal-binding sites, post-translational modifications, and cellular localization. Resources like UniProt \cite{uniprot2025uniprot}, PubMed \cite{white2020pubmed}, and PDB present this information in various forms, including free text entries, abstract summaries, and detailed feature descriptions. Text data is a valuable source for knowledge extraction through natural language processing (NLP) techniques.

\subsection{Data preprocessing}
Transforming raw biological data into usable features for ML models requires several modality-specific preprocessing steps. Raw data are often unstructured or semi-structured and may contain inconsistencies, noise, or missing values — factors that obscure meaningful biological patterns and degrade model performance. As such, preprocessing is essential for cleaning, normalizing, and extracting relevant signals from biological data.
In tumor protein-metal binding, the scarcity of high-quality datasets necessitates the construction of new datasets or the augmentation of existing ones. In these cases, reproducible and well-defined preprocessing workflows are crucial to ensure that the resulting data is suitable for feature extraction and downstream ML model development.
To support this process, Fig. \ref{fig:ftd_fig}B presents a typical data processing workflow for tumor protein-metal binding prediction, including the key steps and tools used to transform raw data into high-quality, structured inputs. We describe the methods for each data modality below.

\textbf{Sequence} data requires three main preprocessing steps to enhance data quality and prepare it for feature extraction. The first step is data filtering and cleaning. Short sequences (e.g., shorter than \rev{45 residues}) are typically removed, as they may represent incomplete segments that do not accurately reflect full proteins \cite{ye2022comprehensive}. Then, highly similar sequences are also removed to reduce redundancy. For example, sequences can be clustered according to a certain identity threshold (typically anywhere between 25\% and 90\%) to retain only one representative from each cluster \cite{cheng2023co,zhang2024metalnet2,shenoy2024m,jamasb2021deep}. MMseqs2 \cite{steinegger2017mmseqs2}, HH-suite \cite{steinegger2019hh}, CD-HIT \cite{fu2012cd}, and Clustal Omega \cite{sievers2014clustal} can streamline the process. The next step in preprocessing is sequence alignment. This can be performed as a pairwise alignment, where a query protein sequence is compared with another to identify regions of similarity. Such alignments can help determine whether the query protein shares homology with known metal-binding proteins, suggesting a potential metal-binding function. More commonly, multiple sequence alignment (MSA) aligns a set of homologous sequences, highlighting conserved residues across the protein family, often indicative of functional or structural importance.
Standard tools for performing sequence alignment include MMseqs2, BLAST \cite{madden2013blast}, and T-Coffee \cite{notredame2000t}. Once aligned, sequences can be annotated with metal-binding information using known metal-binding motifs and structural information extracted from datasets. These annotated sequences provide high-quality inputs for training ML models.

\textbf{Structure} data preprocessing involves four steps. First, in data filtering and cleaning, relevant 3D structures are optimized from datasets using resolution-based filtering. For example, structures with resolutions lower than 3.0 Å are often discarded to ensure reliable geometric details \cite{ye2022comprehensive}. Redundancy reduction is carried out by retaining only one representative structure from clusters of highly similar models, preventing the overrepresentation of a single binding conformation. Redundancy can also be reduced based on sequence identity or through structural comparison using tools such as Foldseek \cite{van2024fast}. Second, structure completion is a crucial preprocessing step that involves cleaning protein structures and imputing missing information. This process involves resolving alternate conformations, adding missing atoms (such as hydrogens) using tools like PDB2PQR \cite{jurrus2018improvements} or Reduce \cite{WORD19991735}, and modeling absent loops with homology modeling tools such as MODELLER \cite{webb2016comparative} or structural predictions from AlphaFold \cite{varadi2024alphafold}. Then, the identification and annotation of metal-binding sites are performed. Metal ions and their coordinating atoms are extracted from PDB or mmCIF files by selecting HETATM records and identifying nearby residues within specific distance thresholds, typically 2.5 to 3.0 Å \cite{ye2022comprehensive,jamasb2021deep}. Furthermore, geometric calculations are performed, such as determining distances, coordination numbers, and bond angles, to accurately characterize the metal coordination environment \cite{xia2024comprehensive}.

\textbf{Protein pocket} shares the following preprocessing steps with structural data: resolution-based filtering, redundancy reduction, identification of related metal ions, and imputation of missing atoms. Beyond these shared steps, pocket-specific preprocessing requires defining the spatial boundaries of the protein pocket and selecting the relevant amino acid residues involved in ligand or metal coordination.

\textbf{Text} data preprocessing in protein-metal interaction studies involves four steps to transform raw textual information into a structured and informative format suitable for computational analysis. Text preprocessing begins with extracting relevant content from datasets such as PubMed or UniProt. The raw text is cleaned to remove irrelevant characters and correct formatting inconsistencies. The next step is tokenization, where continuous text is divided into smaller units called tokens, typically words or subword segments, depending on the granularity required. Normalization techniques, such as lowercase and stemming, are then applied to standardize the text and reduce linguistic variability. Finally, named entity recognition (NER) is employed to identify and classify key biological entities, such as protein names, metal ions, and binding sites \cite{kim2015identifying,cho2019biomedical}.

\subsection{Feature extraction}
Finally, we summarize the unique features that can be extracted from each data modality and present the type of information they carry, along with their contributions to tumor protein research, as shown in Fig. \ref{fig:ftd_fig}C.

\textbf{Sequence-based features} are extracted directly from a protein’s amino acid sequence to determine its metal-binding capabilities. First, the amino acid composition can be represented using one-hot encoding. Second, physicochemical properties, which numerically reflect each amino acid’s biochemical characteristics, offer another valuable feature. For example, each residue can be encoded by properties such as hydrophobicity, polarity, charge, molecular weight, or other experimentally derived indices \cite{xia2024comprehensive,ye2022comprehensive,jamasb2021deep}. This transforms the sequence into a vector of biochemical property profiles. Additionally, evolutionary conservation profiles capture the frequency of amino acids observed in homologous sequences, providing further information. Examples include position-specific scoring matrices (PSSM) \cite{altschul1997gapped} and hidden Markov model matrices \cite{steinegger2019hh}, both of which are used in methods such as GraphBind \rev{GraphBind} \cite{xia2021graphbind}. Finally, deep learning-based sequence embeddings have emerged as a powerful alternative to hand-crafted features. These include word2vec-like approaches such as ProtVec \cite{asgari2015continuous} and more advanced protein language models \cite{heinzinger2025teaching,zhang2025scientific}. These embeddings capture subtle sequence-level semantics and long-range dependencies often missed by traditional encoding techniques.

\textbf{Structure-based features} are derived from the 3D structure of proteins. Secondary structure features refer to the local conformations of a protein’s backbone, which are primarily organized into alpha-helices, beta-sheets, and random coils or loops. DSSP \cite{kabsch1983dictionary} and STRIDE \cite{frishman1995knowledge} are some of the tools used to extract these secondary structure features. These features help define the spatial orientation of metal-ligand residues, thereby identifying potential metal-binding sites. Tertiary structure features capture the complete 3D fold of a protein and include aspects such as the curvature distribution of the protein surface, contact maps that show the distances between atoms or residues in a protein chain, and the spatial convergence of residues that are distant in sequence but come together to form metal-binding pockets. Finally, quaternary structure involves organizing multiple protein subunits into a functional complex. This level of structural organization is essential for modulating protein function as it shapes interfacial interactions and cooperative binding behaviors. Features at this level include the geometry of inter-subunit interfaces, arrangements of cooperative binding sites, conformational dynamics upon ligand binding, and residue interaction networks. These various levels of structural features can be represented as one-dimensional numerical vectors, two-dimensional distance matrices \cite{xia2024comprehensive, jamasb2021deep}, or in the form of a graph using tools like Graphein \cite{jamasb2022graphein}.

\textbf{Protein pocket features} mainly contain geometric descriptors and physicochemical features. Geometric descriptors quantitatively capture the physical characteristics of the pockets. These measurements include pocket volume, surface area, depth, and shape descriptors such as sphericity and curvature. Tools such as Fpocket \cite{le2009fpocket}, CASTp \cite{tian2018castp}, and PocketAnchor \cite{LI2023692} are commonly used to calculate these attributes. Physicochemical features describe the chemical environment within the pocket. For example, the distribution of hydrophobic versus hydrophilic residues, local electrostatic potential, and polarity can be analyzed by examining the amino acid composition of pocket-lining residues. These descriptors can be encoded as numerical vectors, similar to one-hot encoding and physicochemical property encodings used for sequence features. Similarly to structural data, protein pockets can be represented using advanced formats such as graph-based models. In these models, residues that form the binding pocket are treated as nodes, and their spatial proximity or physicochemical interactions are represented as edges. This approach enables the modeling of complex interaction networks that influence metal-binding behavior \cite{jamasb2021deep}.

\textbf{Text features} provide a valuable modality to represent descriptive biological information associated with proteins.
Traditional techniques such as Bag-of-Words (BoW), Term Frequency (TF), and Inverse Document Frequency (IDF) provide a foundational approach for converting text into numerical representations.
BoW captures the frequency of words without considering their order, while TF measures the relative occurrence of each word within a document. IDF adjusts these frequencies by down-weighting common terms and emphasizing rarer ones. The combination of TF and IDF, which is TF-IDF, captures the importance of terms within a document relative to the entire corpus. These simple yet effective methods help highlight relevant vocabulary and term significance in protein-related textual descriptions.
Recent advances in NLP have led to the development of domain-specific language models such as SciBERT \cite{beltagy2019scibert}, BioBERT \cite{lee2020biobert}, and ProtST \cite{xu2023protst}. These models offer valuable initial representations or embeddings of text related to proteins, which can be used to predict their functions, including their ability to bind to metals. Additionally, relationship extraction techniques are employed to identify and structure relationships between entities. For example, a statement like ``protein X binds metal Y at residue Z'' can be transformed into a structured relational triplet \cite{yadav2020relation}. Finally, topic modeling can identify broader themes within the text, capturing functional annotations, biological processes, or experimental conditions relevant to protein-metal interactions.

\section{Multimodal learning for tumor protein-metal binding}
\label{sec:sec5}

Multimodal learning leverages the idea that complex biological entities, such as tumor proteins, can be represented through diverse data modalities, each offering a distinct perspective \citep{liu2025towards}. These sources of information are referred to as modalities, each capturing complementary aspects of the underlying biological phenomenon \citep{baltruvsaitis2018multimodal}. Integrating multiple modalities allows models to compensate for the limitations of individual data modalities and supports a more comprehensive understanding of tumor protein-metal interactions \citep{duan2024deep}.

As discussed in the previous section, four primary data modalities offer insight into different aspects of protein function and behavior. For example, tumor proteins frequently harbor somatic mutations \cite{loeb2000significance}, which can alter their conformation and metal-binding affinity \cite{wang2024amyloid}. Metal binding often induces structural changes that influence downstream biological activity \cite{alfadul2023metal}. Capturing both pre- and post-binding conformational states offers a more nuanced view of metal-induced functional shifts in tumor biology. The tumor microenvironment adds further complexity. Increased protein-protein interactions in this environment can promote complex formation or the emergence of new binding pockets. These dynamics highlight the need for models to integrate structural and contextual information. Textual data, in particular, should go beyond basic annotations to include a disease-specific context. For example, the silencing of MT1G in liver cancer, impairing zinc detoxification, illustrates how gene expression changes can reshape metal-binding behavior \cite{chen2024zinc}. This pathology-specific information enhances the biological relevance of predictive models and supports mechanistic interpretation.

To advance multimodal learning for tumor protein-metal binding, we systematically reviewed existing models developed for this purpose. This section compiles available open-source resources, examines commonly used combinations of data modalities, and reviews representative multimodal learning approaches.

\begin{table}[!t]
\setlength{\tabcolsep}{3pt}
\tiny
\centering
\caption{Key characteristics of machine learning models related to tumor protein–metal binding, as of \rev{May} 2025. Models are sorted by creation date (most recent first) and last update time (most recent first). For the column ``Modality'', ``Seq.'' denotes Sequence and ``Struct.'' denotes Structure. For the column ``Tasks'', ``Localization (Loc.)'' denotes binding site prediction, ``Classification (Class.)'' denotes metal type classification, and ``Identification (Ident.)'' denotes tumor metalloprotein identification. The asterisk (*) in these two columns indicates that the modality or task is not explicitly stated or studied in their paper, but examining the technical content indicates that the particular modality has been used or that the particular task can be a simple extension of the studied task(s), respectively.} 
\label{tab:Table1}
\renewcommand{\arraystretch}{1.92}
\begin{tabularx}{\textwidth}{%
  l c C{1cm} c c c c c c c c c c p{4cm}
  }
 \toprule\midrule
 \multirow{2}{*}{\textbf{Model}} & \textbf{Updated} & \textbf{Prog.} &  \multicolumn{4}{c}{\textbf{Modality}} & \textbf{Interpret-} & \textbf{GitHub} & \textbf{Paper} & \multicolumn{3}{c}{\textbf{Tasks}} & \multirow{2}{*}{\textbf{URL}} \\
 & \textbf{(Created)} & \textbf{language} & Seq. & Struct. & Pocket & Text & \textbf{ability} & \textbf{stars} & \textbf{citations} & Loc. & Class. & Ident. & \\
\midrule
 \multicolumn{14}{c}{\textbf{A. Tumor protein-metal binding models}} \\
\midrule
MetalTrans \cite{zhang2024metaltrans} & 2024 (2024)  & Python & \newcheck & \newcheck & \newcheck & \revcheck\rev{*} & N & 1 & \rev{3} & \newcheck &  & \revcheck\rev{*} & \url{https://github.com/EduardWang/MetalTrans} \\
MetalPrognosis \cite{jia2024metalprognosis} & 2024 (2023) & Python & \newcheck & \newcheck & \newcheck & \revcheck\rev{*} & N & 5 & \rev{2} &  \newcheck &  & \revcheck\rev{*} & \rev{\url{https://github.com/Jrunchang/MetalPrognosis}}  \\
MCCNN \cite{koohi2019predicting} & 2022 (2019) & Python & \newcheck & \newcheck & \newcheck & \revcheck\rev{*} & N & N/A & \rev{68} & \newcheck & & & \rev{\url{https://bitbucket.org/mkoohim/multichannel-cnn}} \\
\midrule
 \multicolumn{14}{c}{\textbf{B. Protein-metal binding models}} \\
 \midrule
MetalNet2 \cite{zhang2024metalnet2} & 2025 (2024) & Python & \newcheck  & & & &  N & \rev{3} &\rev{5} &  \newcheck & \newcheck & & \url{https://github.com/wangchulab/MetalNet2} \\
ESMBind \cite{DAI2025168962} & 2024 (2024) & Python & \newcheck & \newcheck & \newcheck & & Y &\rev{10} &\rev{2} & \newcheck & \newcheck & \revcheck\rev{*} &  \url{https://github.com/Structurebiology-BNL/ESMBind}\\
MetalATTE \cite{zhang2024metalatte} & 2024 (2024) & Python & \newcheck & & & & Y & N/A & 1 & \newcheck & \newcheck & & \url{https://huggingface.co/ChatterjeeLab/MetaLATTE} \\
MIBPred \cite{zhang2024mibpred} & 2024 (2023) & Python & \newcheck &  &  &  & N & 1 & \rev{10} & \revcheck & \newcheck & \newcheck  & \rev{\url{https://github.com/ZhangHongqi215/MIBPred}} \\
Metal3D \cite{durr2023metal3d} & 2024 (2022)& Python & \newcheck &  &  & & \rev{Y} & \rev{39} & \rev{40} & & \newcheck & \revcheck\rev{*} & \url{https://github.com/lcbc-epfl/metal-site-prediction}  \\
GASS-Metal \cite{paiva2022gass} & 2024 (2022) & C++, Python &  & \newcheck & & & Y & 2 & \rev{7} &  \newcheck & \revcheck & \newcheck & \url{https://github.com/sandroizidoro/gassmetal-local}\\
LMetalSite \cite{lmetalsite} & \rev{2024} (2022) & Python & \newcheck &  & & & Y & \rev{19} & \rev{42} & \newcheck & \newcheck & \revcheck\rev{*} & \url{https://github.com/biomed-AI/LMetalSite} \\
M-Ionic \cite{shenoy2024m} & 2023 (2022) & Python & &  \newcheck & & & \rev{Y} & 4 & \rev{12} &  \newcheck &  \newcheck &  \newcheck & \url{https://github.com/TeamSundar/m-ionic}\\
MIB2 \cite{lu2022mib2} & 2022 (2022) & N/A & \newcheck & \newcheck & & & N & N/A & \rev{85} & \newcheck & \newcheck & \revcheck\rev{*} & \url{http://bioinfo.cmu.edu.tw/MIB2/}\\
MEBIPRED \cite{aptekmann2022mebipred} & \rev{2025} (2021) & N/A & \newcheck &  & & & N & N/A & \rev{28} & \newcheck &\revcheck & \newcheck & \url{https://services.bromberglab.org/mebipred/home} \\
MetalNet \cite{cheng2023co} & 2024 (2021) & Python & \newcheck & & & & \rev{Y} & \rev{19} & \rev{30} & \newcheck & \newcheck & \revcheck\rev{*} & \url{https://github.com/wangchulab/MetalNet}\\ 
MetalSiteHunter \cite{mohamadi2022ensemble} & 2022 (2021) & Python & &  \newcheck & \newcheck & & Y & \rev{5} & 13 & \newcheck & \newcheck & \newcheck & \url{https://github.com/ClinicalAI/metal-site-hunter} \\
BioMetAll \cite{sanchez2020biometall} & 2024 (2020) & Python &  & \newcheck & & & \rev{Y} & \rev{12} & \rev{42} & \newcheck & \newcheck & \newcheck & \url{https://github.com/insilichem/biometall}\\
MIonSite \cite{qiao2019mionsite} & 2018 (2018)& Perl & \newcheck & & & & N & \rev{5} & 22 & \newcheck & \newcheck & \revcheck\rev{*} & \url{https://github.com/LiangQiaoGu/MIonSite}\\
\midrule
\multicolumn{14}{c}{\textbf{C. Protein foundation models}} \\
\midrule
Oneport \cite{floge2024oneprot} & 2025 (2024)& Python & \newcheck  &  \newcheck & \newcheck & \newcheck & N & 13 & 1 & \newcheck & \newcheck & \newcheck & \url{https://github.com/klemens-floege/oneprot}\\
ProteinChat \cite{huo2024multi} & 2025 (2024)& Python & \newcheck  &  & & \newcheck & \rev{Y} & \rev{10} & 3 & \newcheck & \newcheck & \newcheck & \url{https://github.com/mignonjia/ProteinChat}\\
ProTrek \cite{su2024protrek} & \rev{2025} (2024)& Python & \newcheck  & \newcheck & & \newcheck & N & \rev{101} & \rev{13} & \newcheck & \newcheck & \newcheck & \url{https://github.com/westlake-repl/ProTrek}\\
ProteinAligner \cite{zhang2024proteinaligner} & \rev{2025} (2024) & Python & \newcheck & \newcheck & & \newcheck  & N & \rev{11} & 0 & & & \newcheck &\url{https://github.com/Alexiland/ProteinAligner}\\
\midrule
 \bottomrule
\end{tabularx}
\end{table}

\subsection{Research advances and open resources}
\label{sec:sec3}

We review existing literature to identify studies that explore the application of ML in tumor protein-metal binding. Table \ref{tab:Table1} summarizes the following key characteristics of these resources: \textbf{Prog. language} indicates the programming language used; \textbf{Modality} lists the data modalities used; \textbf{Interpretability} specifies whether the model incorporates interpretability analyses beyond basic data validation or model comparison; \textbf{Creation and last update time} reflects the recency and maintenance status, providing insight into its ongoing usability and relevance; \textbf{GitHub stars} is a proxy of reputation and influence within the developer and research communities. This criterion is included because most relevant software is hosted and maintained on GitHub \cite{escamilla2022rise}; \textbf{Paper citations} serve as a proxy of the academic impact and visibility of the work, \add{where the citation count is based on Google Scholar}; \textbf{Tasks} summarize the scope and functional capabilities of each work; and \textbf{URL} provides access to the public code repository or official website. When both are available, only the code repository is listed.

We first study three recent models designed for tumor protein–metal binding in Table \ref{tab:Table1}A: multichannel convolutional neural network (MCCNN) \cite{koohi2019predicting}, MetalPrognosis \cite{jia2024metalprognosis}, and MetalTrans \cite{zhang2024metaltrans}. 
MCCNN is the first deep learning model designed to predict disease-related mutations at metal-binding sites in metalloproteins, providing a novel platform for exploring the role of metal-binding disruptions in human disease. MCCNN integrates three types of data: protein sequence, 3D structural data of binding sites (including protein structure and pocket features), and metadata including categorical and textual data. Therefore, MCCNN can be considered as integrating four modalities, sequence, structure, protein pocket, and text data, from four datasets of protein mutations (Tables \ref{tab:CancerDatabases} and \ref{tab:protein-metal binding datasets}): MetalPDB \cite{putignano2018metalpdb}, CancerResource2 \cite{gohlke2016cancerresource}, ClinVar \cite{landrum2016clinvar}, and Uniprot Humsavar \cite{wu2006universal}.
MetalPrognosis \cite{jia2024metalprognosis} and MetalTrans \cite{zhang2024metaltrans} are developed to further enhance predictive performance in tumor protein–metal binding. Both models were built on the MCCNN dataset, introducing architectural improvements to enhance accuracy and efficiency.
MetalPrognosis eliminates the need for manual feature extraction by using sliding window sequences as input. These sequences are processed through pre-trained protein language models to extract rich semantic representations, which are then fed into a convolutional neural network to capture complex, high-level features.
MetalTrans integrates multiple features via concatenation within a transformer-based framework, enabling comprehensive and context-aware feature extraction across modalities.

To the best of our knowledge, MCCNN, MetalPrognosis, and MetalTrans are the only ML models designed for tumor protein-metal binding prediction. Since tumor proteins are not structurally distinct from general proteins, the learning mechanisms in ML models are transferable across related domains. Therefore, we expand our analysis scope to include general protein-metal binding models and open their potential applicability in tumor protein research.
As shown in Table \ref{tab:Table1}B, we have collected 15 recent protein-metal binding prediction models that can be adapted to tumor-specific tasks through training or fine-tuning with tumor protein. For example, LMetalSite \cite{lmetalsite} and M-Ionic \cite{shenoy2024m} can serve as valuable baselines for evaluating the performance of MetalPrognosis. 
Besides protein-metal binding prediction models, recent advances in foundation models have introduced pre-training on large-scale general datasets, followed by fine-tuning for specific downstream tasks. We include four representative protein foundation models in Table \ref{tab:Table1}C that can support tumor protein-metal binding tasks. These models utilize powerful protein language models to extract rich contextual representations, offering promising avenues to enhance the diversity and depth of tumor protein–metal binding prediction.

\add{We can draw the following insights from Table 3: 
\begin{itemize}
    \item Research on predictive models for tumor protein-metal binding is still nascent, with the classical MCCNN model currently prevailing. This highlights the urgent need for methodological advancements in this field. In contrast, the field of protein-metal interactions has seen a proliferation of diverse models. Notably, protein foundation models based on large language models have garnered significant academic attention (e.g., high-star GitHub projects), offering new technological directions.
    \item The openness of the Python ecosystem has lowered model development barriers, benefiting the growth of the research community. Current mainstream models integrate multimodal data (sequence, structure, pocket, and text), but biological authenticity modeling can be enhanced. Future models can incorporate multiprotein interaction networks and dynamic microenvironments, and leverage tumor multi-omics architectures to integrate cross-scale information, such as chemical microenvironments and radiomics.
    \item Existing models primarily focus on binding site prediction. Future work can expand to predicting specific metal binding types based on site features and establishing bidirectional prediction systems for metal ligand-protein structures, which will drive the integration of metal-binding mechanism research and tumor-targeted therapy.
\end{itemize}
}
\subsection{Modality combinations}
\label{sub:mm2}
As shown in Table \ref{tab:Table1}, several unimodal methods have been proposed for protein-metal binding prediction using sequence \cite{zhang2024metalatte,shenoy2024m,cheng2023co,aptekmann2022mebipred} or structural data \cite{buchanan2019metal,paiva2022gass,sanchez2020biometall}. However, relying on a single modality limits the ability to capture the full complexity of tumor protein-metal interactions. Integrating multiple data modalities, on the other hand, can improve predictive accuracy and model interpretability, thereby supporting more comprehensive computational analyses. Table \ref{tab:Table1} presents various modality combinations for protein-metal binding prediction and the related protein foundation model that can enhance and enrich tumor protein-metal binding prediction models.

\textbf{Sequence and structure} each provide valuable but distinct insights into protein-metal interactions. Sequence data alone may overlook the spatial relationships and binding-site conformations that are critical for accurate metal-binding prediction. Conversely, structural data is often constrained by the limited availability of experimentally validated protein structures, especially for novel or mutated tumor proteins. Greener et al. primarily utilized the data of the metalloprotein sequence and supplemented it with structural insights derived from the ``periodic table'' of protein structures, based on Taylor's work \cite{greener2018design}. This approach enabled the generation of synthetic sequences corresponding to new protein topologies.

\textbf{Sequence and text} complement each other by allowing the models to access both primary sequence data and biological contextual information. For example, ProteinChat \cite{huo2024multi} is trained on over 1.5 million carefully curated triplets (protein, text, answer) from the Swiss-Prot dataset to capture protein knowledge, mimicking expert reasoning.

\textbf{Sequence, structure, and text} can be effectively combined to address annotation gaps in existing protein databases. Currently, approximately 30\% of the proteins in UniProt remain unannotated. ProTrek \cite{su2024protrek} addresses this gap by leveraging sequence, structure, and text data to predict and annotate the functions of previously unidentified proteins. Natural language models can further enhance this task by extracting insights from rich textual descriptions while incorporating sequence and structural information. For example, ProteinAligner \cite{zhang2024proteinaligner} adopts this approach to identify post-translational modifications, capture protein dynamics under different conditions, and map interaction networks directly from the literature.

\textbf{Sequence, structure, and protein pocket} provide a detailed understanding of site-specific information. Trained with structures predicted by AlphaFold \cite{jumper2021highly}, ESMBind \cite{DAI2025168962} addresses the challenge of accurately determining the 3D coordinates of metal ions in protein structures, outperforming models such as LMetalSite and M-Ionic, which do not incorporate protein pocket information. ESMBind enables deep learning-based metal-binding prediction at multiple levels, including sequence-level, residue-level, and atomic-level modeling.

\textbf{Sequence, structure, protein pocket, and text} represent a comprehensive integration of the previously discussed data modalities. This multimodal combination enables the capture of both low-level molecular features and high-level biological context. The Oneport model \cite{floge2024oneprot} exemplifies this approach by generating unified protein representations that support accurate cross-modal retrieval even between modality pairs not explicitly trained together, demonstrating strong generalization and biological relevance. In the context of tumor protein–metal binding, such multimodal combinations are equally applicable. For example, MCCNN effectively integrates multimodal data for targeted prediction tasks in this domain.

\rev{\subsection{Learning from multimodal data}
\label{sub:mm3}
To learn from multimodal data, we need a fusion scheme and modality representation learning methods. Multimodal fusion schemes can be categorized into early (data-level), intermediate (feature-level), late (decision-level), or hybrid fusion (mix of the previous three schemes) \citep{meng2020survey,zhang2021deep}. In the following, we discuss various approaches to learn feature representations from multimodal data \citep{10.1145/3649447}.}

\textbf{Classical ML models}, such as decision trees, support vector machines, and random forests, have been applied to multimodal tasks with concatenated features from different modalities \cite{dhakal2022artificial}. While these models are interpretable and computationally efficient, they are limited in learning complex relationships or representations from high-dimensional and unstructured data.

\textbf{Deep learning models} can effectively capture complex, nonlinear relationships within and across data modalities. 
Convolutional neural networks (CNNs) can capture local patterns in structured inputs. In the context of tumor protein–metal binding, CNNs can be adapted to learn protein sequences, 3D structures, and descriptive textual information separately or in combination \cite{kha2022identifying,igashov2021vorocnn,alachram2021text}.
Transformer models \citep{vaswani2017attention} use self-attention mechanisms to capture relationships between input features across modalities. ProteinAligner \cite{zhang2024proteinaligner} utilizes eight transformer layers to process sequence, structural, and textual data. MetaLATTE \cite{zhang2024metalatte} further introduces a position-sensitive attention mechanism to predict metal specificity, including zinc, lead, and mercury.
Graph neural networks are particularly effective for modeling biological data with intrinsic topological structures, such as molecular interactions and protein conformations. For example, graph convolutional networks (GCNs) have been used to predict residue-level binding affinity between yellow fluorescent proteins and rare earth elements \citep{tvedt2024predicting}, and extract complementary features from multi-omics data (mRNA expression, DNA methylation, and microRNA profiles) for disease classification tasks \citep{wang2021mogonet}. \add{Deep learning relies heavily on large quantities of high-quality data. However, annotated data on tumor protein-metal binding, especially for mutated samples, are scarce. For instance, despite the availability of tumor-related datasets, MetalPrognosis \cite{jia2024metalprognosis} and MetalTrans \cite{zhang2024metaltrans} still need to leverage cross-species or general protein data for transfer learning.}

\textbf{Generative models} perform feature learning with reconstruction capabilities and are increasingly used for multimodal representation learning and synthetic data generation. Many generative models are built on the encoder–decoder architecture, which transforms multimodal input into a shared latent space and reconstructs it through a decoder. For example, ESBind \cite{DAI2025168962} uses such architecture to learn from sequence and structural data for predicting protein–metal binding. Multimodal deep autoencoder \citep{Wang2020ANA} learns joint representations by fusing similarities from multiple modalities to support downstream tasks like drug–target interaction. A more advanced variant, Variational Autoencoders \citep{wang2024advancing}, incorporates both an inference mechanism and a generative process to model the underlying data distribution and produce compact latent representations of sequence and structural data \cite{greener2018design}. In addition to representation learning, generative models can address data scarcity by producing synthetic data, such as generating 3D structures from protein sequences \citep{cui2019predicting} and creating descriptive text to supplement incomplete biological annotations. \add{However, the biological validity of such generated results can be questionable. For example, the predicted metal coordination geometry may not match the actual bond lengths and coordination numbers \cite{DAI2025168962}. Moreover, generation based on known metal-binding protein information may limit their generalization ability for novel metals (such as rare earth elements) or rare tumor mutations \cite{DAI2025168962}.}

\rev{While these models are not yet widely used for tumor protein–metal binding, their application in related domains demonstrates their potential to enhance predictive performance and uncover biological insights in this area.}

\section{Interpretability in tumor protein-metal binding models}
\label{sec:sec6}

Interpretability in ML refers to model transparency, ease of user understanding, and the capacity to foster trust \citep{nasarian2024designing}. It is commonly defined as the extent to which the reasoning behind model predictions can be clearly explained \citep{biran2017explanation}. In tumor protein-metal binding, interpretability is particularly important — not only to understand how predictions are made but also to ensure that these predictions are meaningful and actionable for biologists and clinicians \citep{imrie2023multiple}. 

As shown in Table \ref{tab:Table1}, current models place limited emphasis on interpretability, particularly in the context of tumor protein–metal binding. \rev{In the following, we first examine the importance of interpretability and its specific relevance to tumor protein–metal binding, and then explore  interpretable ML methods for computational biology and summarize their key limitations.}

\subsection{Importance of interpretability}

Complex ML models, particularly deep learning models, are often considered ``black boxes'' because of their complexity and lack of interpretability \citep{rudin2019stop}. While these models can achieve high predictive accuracy, their opacity poses significant challenges in understanding the underlying biological mechanisms of protein-metal binding sites and limits their utility in scientific discovery \citep{chen2024applying}. The bias embedded in these models can lead to unfair or misleading results, which is especially concerning in biomedical research \citep{obermeyer2019dissecting}. Additionally, deep learning frequently yields inconsistent results in different datasets or experimental conditions, further complicating validation and reproducibility efforts \citep{hutson2018artificial}.

\rev{Interpretable ML} methods can address these issues by demystifying the modal decision-making process, enabling researchers to verify the biological mechanisms captured by the models and establish a scientific basis for predictions \citep{chen2024applying}. These methods also help identify and reduce bias, ensuring that the models are fair and reliable \citep{obermeyer2019dissecting}. Furthermore, by providing clear and understandable explanations of the behavior of the model, interpretable ML methods improve reproducibility and facilitate the verification of prediction results \citep{ribeiro2016should}. An interpretable model can promote new scientific discoveries by offering detailed insights into the underlying processes modeled, thereby advancing scientific progress \citep{avsec2021base,ma2018using,vig2020bertology,barnett2021case}.

Enhancing \rev{ML model interpretability} can help scientists gain deeper insight into the interactions between metal ions and tumor proteins, including binding sites, binding modes, and downstream biological effects. Interpretability techniques help highlight key structural features and amino acid residues, offering precise targets for the design and optimization of anti-tumor drugs. By predicting potential binding relationships between novel tumor proteins and metal ions and by evaluating the reliability of these predictions, interpretable models support more informed decision-making. As a result, they can potentially accelerate drug development while reducing experimental costs.

\add{Most ML-based studies in tumor protein-metal binding or broader protein-metal binding focus primarily on predictive capability, with limited interpretability studies. Consequently, the mechanisms behind performance improvement or differences are often unclear. To better understand these challenges, we next examine existing interpretable ML methods and their potential to enhance biological understanding and domain-specific reliability (Fig. \ref{fig:Fig.1}C).}

\rev{\subsection{Interpretable ML methods}

Following two recent reviews \citep{nasarian2024designing,chen2024applying}, interpretable ML methods for computational biology can be categorized into two approaches: designing inherently interpretable models (i.e., by-design models) and performing post hoc analysis. 

In tumor protein-metal binding research, the focus was on certain specific amino acids, such as histidine (His) and cysteine (Cys), because these amino acids contain side-chain groups that can coordinate with metal ions \cite{farkas2016metal,martin2024complexes}. By conducting site-directed mutagenesis experiments to alter these amino acids, changes can be observed in metal binding capabilities to identify binding sites. Meanwhile, in biological experiments, techniques such as X-ray crystallography and nuclear magnetic resonance (NMR) can be utilized to determine the three-dimensional structures of proteins for analyzing the structural environments of metal binding sites \cite{handing2018characterizing,jensen2005metal}. For instance, some metal binding sites may be located near hydrophobic pockets of proteins or within specific domains. 

Using by-design models or performing post hoc feature importance analysis on ``black-box'' models can facilitate the investigation of whether prediction models can accurately capture the complex binding structures of specific amino acids and metals, as well as their relationships to predicted binding sites.%

\subsubsection{By-design models}
Transparent models, such as linear models and decision trees, have simple structures that make the relationships between inputs and outputs easy to understand. These models are inherently interpretable by design. For example, logistic regression and linear support vector machines (both are linear classifiers) have been applied to classify the chemical stoichiometry of virus-like particles (VLPs) \cite{zhang2025classifying}. The learned model weights reveal which parts of the protein sequences contribute more to the binary classification of 60-mer vs 180-mer. The results also show that key amino acids at the ends of $\beta$-sheets (such as valine and isoleucine) may influence VLP assembly more by stabilizing subunit interactions. This application of interpretable linear models can potentially reduce the need for costly lab testing and guide the rational design of VLPs.

Biologically inspired models incorporate known biological concepts into decision-making \cite{cheng2023co,sanchez2020biometall,nilmeier2013rapid}. For example, GASS-Metal constructs similarity matrices from homologous proteins with the same Enzyme Commission number, rather than relying on general-purpose ones like Blosum62 \cite{nilmeier2013rapid}. This approach reveals how metal-binding functionality is maintained, e.g., through conservative substitutions that preserve key physicochemical properties, even when there are significant sequence differences. This aligns with the hypothesis of ``function-driven structure-sequence coevolution'' and enhances the trustworthiness of the model.

To capture spatial information and functional interactions in proteins, attention mechanisms have been increasingly used. Attention-based models can learn to identify specific residues or regions that are most relevant to the prediction task. For example, MetalATTE incorporates position-sensitive attention pooling and rotary position embeddings to enhance identification of critical binding sites \cite{zhang2024metalatte}. Its attention weights highlighted cysteine residues and neighboring amino acids (e.g., histidine, aspartate) as key contributors to heavy metal binding. Furthermore, this interpretability supports biological design. In one case, substituting cysteine residues with alanine at key binding sites led to predicted and experimentally validated loss of function (i.e., non-binding), consistent with the role of cysteine clusters in metal chelation \cite{bofill2009independent}.

\subsubsection{Post hoc analysis}
While by-design models offer built-in interpretability through their structure or biological grounding, post hoc methods aim to explain the predictions of already-trained models, especially when using complex or black-box architectures.

Post hoc interpretability methods can reveal which features or regions a trained model relies on for its predictions. Gradient-based and perturbation-based approaches, such as integrated gradients, Grad-CAM, SHAP, and LIME, are commonly used for this purpose \cite{shrikumar2017learning,selvaraju2020grad,lundberg2017unified,ribeiro2016should}. These methods can help identify influential sequence features or structural positions after the model has been trained.

For example, the hierarchical graph transformer with contrastive learning applies Grad-CAM to project deep model decisions onto the three-dimensional structures of proteins \cite{gu2023hierarchical}. By visualizing residue-level contributions, it identifies functional sites and overcomes the ``homogenized'' representations often seen in traditional approaches. In one case, when predicting the function of a protein (e.g., A0A3P7DWR6), Grad-CAM highlighted DNA binding sites that were highly conserved in homologous proteins (e.g., 5H1C), suggesting these residues are evolutionarily constrained and functionally important.

Similarly, Positional SHAP introduces position-sensitive SHAP values to interpret models at a fine-grained level \cite{dickinson2022positional}. This approach not only recovers known biological motifs (such as MHC anchor sites) but also uncovers non-trivial interactions between sequence positions, such as those involving electrostatic effects or spatial dependencies.

\subsubsection{Interpretability in multimodal integration}
Multimodal information integration can improve model performance, but it also raises new questions about the relative contributions and interactions between different modalities. For instance, is it always beneficial to integrate protein sequence and structural information? How can their respective influences be measured? In some tumor proteins with similar sequences but different structures, structure may dominate the identification of metal-binding sites. Conversely, for proteins with similar structures but low sequence conservation, sequence information might be more informative. These hypotheses require systematic testing, and ablation studies provide a principled approach to exploring them.

Ablation experiments reveal potential causal relationships between inputs and outputs by selectively removing specific features or modalities \cite{shenoy2024m,DAI2025168962}. ESMBind validates the necessity of integrating sequence and structure by comparing its full model with variants using only sequence (ESM-2) or only structure (ESM-IF) \cite{DAI2025168962}. Its findings also suggest that experimental validation of light metal-binding proteins should prioritize structural dynamics, whereas transition metal binding may rely more on sequence conservation.

From a modeling perspective, predictions about metal-binding sites often rely on learned discriminative criteria, such as statistical thresholds for residue co-occurrence or patterns in local structural geometry. Yet, are these criteria consistent with the biochemical principles underlying metal coordination? Integrating biological prior knowledge into model decision-making, either during training or in model architecture, can help align computational predictions with established biochemical mechanisms.

}

\add{

\subsection{Limitations of interpretable ML methods}
Although recent years have seen progress in interpretable ML methods for protein research, these approaches continue to face fundamental challenges. Addressing these limitations is key to ensuring that interpretable ML becomes a reliable tool in real-world biological and clinical applications.

From the perspective of method evaluation, the absence of ``ground truth explanations'' poses a significant obstacle to assessing the biological accuracy and relevance of interpretability outcomes \cite{zhou2021evaluating}. More importantly, the current lack of a standardized evaluation framework makes it difficult to systematically assess and compare interpretable ML methods across diverse tasks and domains \cite{zhang2019towards, carvalho2019machine}.

Interpretable methods themselves also have inherent limitations. Feature importance analyses can highlight statistically significant features; however, these often lack clear biological meaning or relevance \cite{huynh2012statistical}. Moreover, the presentation formats of interpretability results, such as SHAP values, often diverge from the cognitive expectations of domain experts, impeding clinical translation and practical utility \cite{salih2025perspective}. In addition, instability observed in methods like SHAP and LIME raises concerns about their reliability \cite{chen2024applying}. Causal inference approaches, such as ablation studies, can evaluate the necessity of specific features but generally fall short of uncovering mechanistic biological insights at the molecular level \cite{kelly2022review}.

Emerging multimodal foundation models present new challenges, particularly in their tendency to favor textual representations while underutilizing essential features from other modalities such as protein structure and physicochemical properties \cite{liang2021towards, queen2024procyon}. This bias may risk distorting the interpretability of such models in biological applications.

As model architectures grow more complex, relying on a single interpretability method has become limited in capturing the full decision-making process \cite{toussaint2024explainable}. Effectively combining multiple interpretability strategies while preserving their individual strengths remains an open challenge. Furthermore, a major concern in current interpretability research is the selective emphasis on well-performing, locally interpretable decisions, with much less effort devoted to explaining where models generalize poorly \cite{chen2024applying}. This selective transparency hinders robust evaluation and obstructs the translation of ML models from theoretical promise to clinical or experimental practice.

}

\section{Perspectives and future directions}
\label{sec:sec7}

Finally, we summarize key challenges that hinder progress in interpretable multimodal learning for tumor protein–metal binding and discuss promising directions based on insights from previous sections. Beyond these solutions, we highlight two emerging research directions that extend current efforts and have the potential to further advance this domain.

\subsection{Challenges and opportunities in prediction}
Several technical and practical barriers remain unresolved in this domain. First, the scarcity of high-quality, tumor-specific datasets limits the training and generalization capabilities of ML models. Second, integrating diverse data modalities, such as protein sequences, structures, binding pockets, and textual annotations, introduces complexity, as each carries distinct formats and levels of biological insight. Developing robust fusion strategies that preserve meaningful signals across modalities remains a significant challenge. Lastly, the ``black-box'' nature of complex ML models remains a barrier to transparency, particularly in biomedical research where interpretability is essential for credibility and clinical translation. 

To address these challenges, we explore the following promising directions:
\begin{itemize}
    \item \textbf{Integrating cancer-related and protein-metal binding datasets}: Combining existing tumor protein datasets (e.g., TCGA) with protein-metal binding datasets (e.g., PDB, MetalPDB) can help address data scarcity and generate more diverse, informative training datasets for ML model development. 
    \item \textbf{Adapting existing models for tumor-specific prediction}: Existing protein–metal binding models (e.g., LMetalSite, M-Ionic) can be adapted to tumor-specific settings to support targeted prediction tasks. Additionally, pre-trained protein foundation models can be fine-tuned and integrated for tumor-relevant applications.
    \item \textbf{Leveraging multimodal learning with standardized data workflows}: Standardizing well-defined data processing and feature extraction workflows ensures high-quality inputs for ML models. Pairing standardized workflows with appropriate multimodal fusion strategies and ML models can maximize the complementary strengths of diverse data modalities, leading to improved model performance and robustness.
    \item \textbf{Improving model interpretability}: Incorporating inherently interpretable ML models (e.g., decision trees, linear models) \rev{and other by-design interpretable ML models and applying post hoc explanation methods (e.g., LIME, SHAP), while addressing their limitations,} can help uncover biologically meaningful insights and promote trust in ML-driven predictions.
\end{itemize}

\add{Together, these strategies provide a foundation for improving prediction accuracy, transparency, and translational potential. In addition to these strategies, biological complexity presents further challenges for accurately predicting tumor protein–metal binding. The tumor microenvironment is a dynamic and heterogeneous system in which factors such as metal ion concentrations and oxidative stress can influence both the likelihood and outcome of metal-binding events. One way to address this complexity is through multi-omics integration models that incorporate metabolomics-derived features, such as metal ion concentrations and oxidative stress biomarkers, together with protein structural data to help elucidate the dynamic distribution of metals within the tumor microenvironment and their impact on protein conformation \cite{zhou2013proteomics,zhou2022metalloproteomics}. Incorporating such context-aware biological information may support more realistic modeling of metal–protein interactions in vivo and motivate future integration of spatiotemporal data and simulation-based approaches, ultimately bridging the gap between predictive modeling and real-world biological systems.}

Beyond these practical strategies, we highlight two emerging directions that could transform our understanding of tumor protein–metal interactions: integrating protein–protein interaction (PPI) data can provide structural context for metal-binding events, while modeling post-binding structure changes in tumor proteins offers new insights into how metal coordination reshapes structure and function. These directions represent promising extensions of current efforts, with the potential to enhance both mechanistic understanding and therapeutic development.

\subsection{Integrating protein–protein interaction networks for contextual prediction}

Protein-protein interaction (PPI) networks are essential in cellular function, and tumor proteins often exert their effects through these networks. 
Incorporating PPI information can provide valuable functional context for understanding the biological significance and functional impact of metal-binding events. 
For example, the organoarsenic drug Darinaparsin (ZIO-101) has demonstrated anticancer activity across various tumor cell lines. Studies have shown that ZIO-101 can bind to histone H3.3, affecting the function of histone H3.3 in the cell nucleus \cite{xu2019s}. PPI analysis reveals that histone H3.3 closely interacts with histone deacetylase 1 (HDAC1). HDAC1 can regulate the level of histone acetylation. This process further affects the expression of cyclin-dependent kinase inhibitor 1A (CDKN1A). Ultimately, it leads to the overexpression of genes related to tumor necrosis factor, which induces apoptosis and thus inhibits tumor growth. Focusing solely on the binding of histone H3.3 to arsenic without considering its interaction with HDAC1 would fail to fully grasp the anticancer mechanism of arsenic compounds and could overlook potential therapeutic strategies targeting HDAC1 \cite{xu2019s}.
These examples highlight that PPI often influence the formation, stability, and function of metal-binding sites. PPI data can enhance model accuracy by providing biologically relevant context \cite{zhang2024proteinaligner}. However, current prediction models make limited use of PPI information. Therefore, we encourage future research to integrate PPI networks to improve the functional relevance and interpretability of tumor protein–metal binding predictions.

Recent advances in bioinformatics have provided a solid foundation for this integration. Public datasets, such as STRING \cite{szklarczyk2025string} and PINA \cite{du2021pina}, have accumulated extensive PPI data that can be mined to extract interaction networks relevant to tumor protein-metal binding functions. At the same time, computational methods such as molecular docking and molecular dynamics simulations enable researchers to model protein-protein interactions and complexes. They can also investigate how metal binding sites change in these cases \cite{sankaranarayanan2024molecular}. In addition, experimental techniques such as yeast two-hybrid screening and coimmunoprecipitation can validate predicted interactions, providing strong experimental support for ML models \cite{bruckner2009yeast,tang2018analysis}. User-friendly PPI network analysis platforms improve accessibility for researchers with limited programming expertise, facilitating broader adoption of PPI-informed modeling strategies \cite{sanz2012tools}. As such, integrating PPI data can significantly enhance the biological fidelity and interpretability of tumor metal-binding predictions.

\subsection{Modeling post-binding structural changes in tumor protein–metal complexes}
The 3D conformation of a protein underpins its biological function, including catalysis, signal transduction, and structural support. When binding to a metal ion, proteins often undergo structural changes that alter their function. This also holds for tumor proteins, where metal binding may disrupt or modulate oncogenic processes. Metal-based drugs, such as platinum-based drugs, exploit this mechanism to alter tumor protein structures and induce cell death. 

To model this process computationally, we propose using existing protein–metal binding predictive methods and extending them to estimate the structural changes in tumor proteins after metal binding. These models can serve as valuable tools for assessing therapeutic potential, accelerating drug development, and overcoming the limitations of experimental structural determination.
Existing models for studying protein-metal interactions operate within a ternary framework composed of three key components: the protein, its binding site(s), and the associated metal ions. For example, models using protein and binding site data as inputs can predict the types of metals that may bind. Conversely, models using protein and metal information can identify whether a protein functions as a metalloprotein and locate its potential binding sites. Through these modular configurations, models can infer conformational changes even with incomplete input data, enabling structure prediction under conditions where structural data are missing or limited \cite{zhang2024proteinaligner}. 

\begin{figure}[!t] %
    \centering
    \includegraphics[width=1\textwidth]{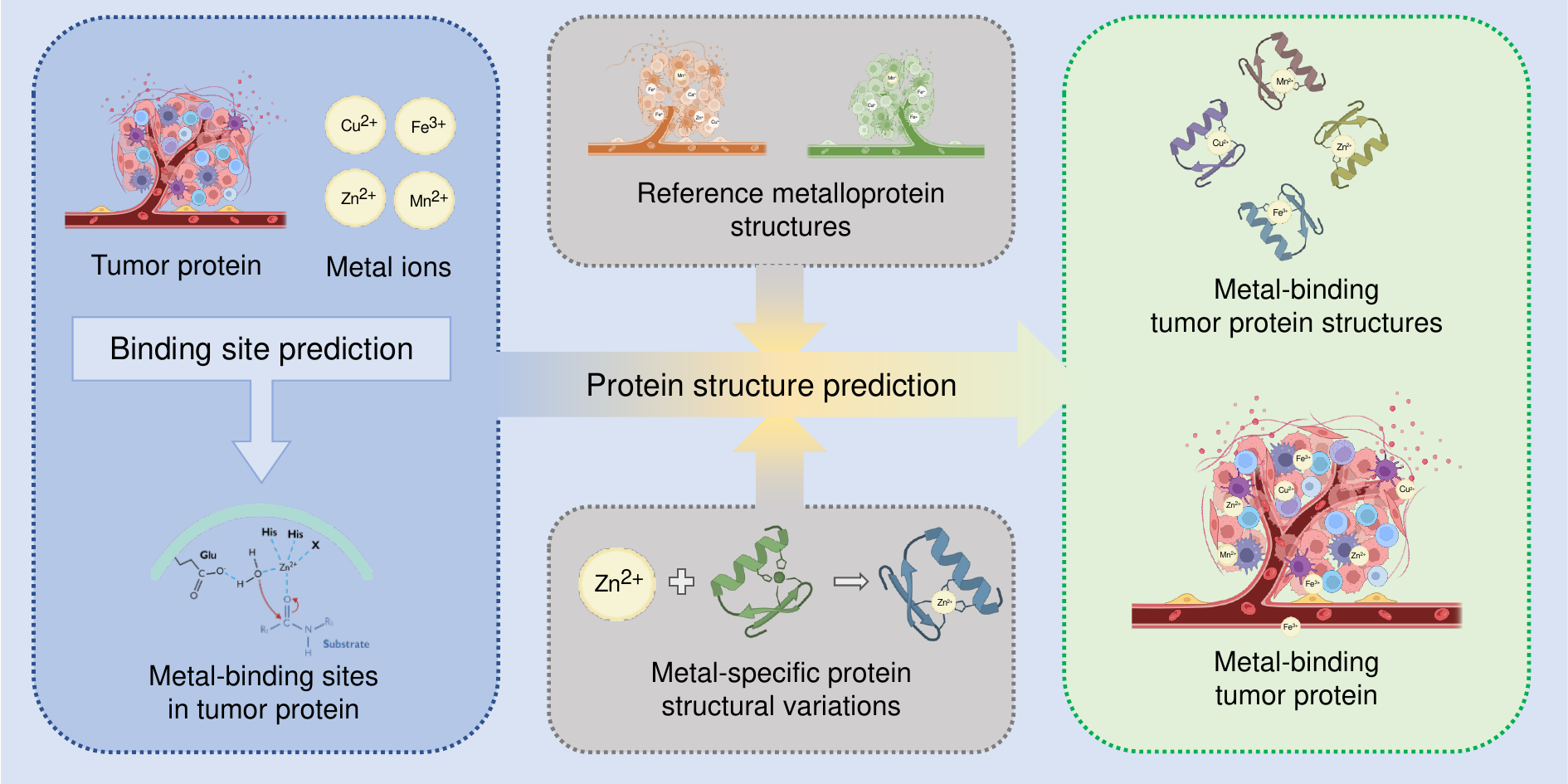} %
    \caption{Post-binding structural prediction in tumor protein-metal complexes. First, binding sites are predicted using established tumor protein–metal binding models. Next, the post-binding structure is modeled using the protein, metal ion, and predicted binding sites as input. To enhance structural accuracy, the framework incorporates reference metalloprotein structures and known metal-specific structural variations. This approach aims to generate biologically meaningful predictions of metalloprotein conformations associated with tumors.}
    \label{fig:Fig.3} %
\end{figure}

Based on recent studies \cite{dai2025predicting,DAI2025168962}, we propose a framework for post-binding structural prediction in tumor protein-metal complexes, as shown in Fig. \ref{fig:Fig.3}. The framework operates in two stages: first, it predicts binding sites using inputs from tumor proteins and metal ions; then, it utilizes the protein, metal, and predicted binding sites to model the post-binding structure. To enhance prediction accuracy, the framework incorporates prior knowledge from known metalloprotein structures and metal-specific structural effects. Additionally, domain-specific constraints, such as chirality and geometric coordination rules, are applied to ensure biological plausibility. This framework facilitates the predictive modeling of metal-bound tumor protein structures, providing a pathway toward structure-guided drug design targeting metal–protein interactions in cancer.

\section{Conclusion}
\label{sec:sec8} 
\rev{
Interpretable multimodal learning presents a promising avenue for enhancing the efficiency and transparency of tumor protein–metal binding research. In this paper, we have outlined a perspective shaped by insights from existing protein–metal binding studies, encompassing diverse data modalities, multimodal learning strategies, and model interpretability. We highlighted the importance of standardizing data acquisition, optimizing multimodal integration, and developing interpretable machine learning models. Beyond these core areas, we proposed two emerging directions for advancing future research: integrating protein–protein interaction data to provide functional context, and modeling structural changes in tumor proteins following metal binding to inform therapeutic development. We believe that realizing this vision will require close collaboration between clinicians, biologists, and machine learning researchers, establishing a strong foundation for impactful and interdisciplinary progress in the field.
}

\section{Declaration of competing interests}
The authors declare that they have no known competing financial interests or personal relationships that could have appeared to influence the work reported in this paper.

\section{Acknowledgments}
This work is supported in part by the National Natural Science Foundation of China (No. 32301200), Fundamental Research Program of Shanxi Province (No.202403021221020), Shanxi Scholarship Council of China (No. 2023-006), and Shanxi Provincial Basic Research Program (No. 202303021222147).

\section{Declaration of generative AI and AI-assisted technologies in the writing process}

During the preparation of this work, the authors used ChatGPT and Kimi in order to improve language and readability. After using these tools/services, the authors reviewed and edited the content as needed and take full responsibility for the content of the publication.

\bibliography{output}

\begin{appendices}

\counterwithin{figure}{section}
\counterwithin{table}{section}
\counterwithin{equation}{section}
\renewcommand\thefigure{\thesection\arabic{figure}}
\renewcommand\thetable{\thesection\arabic{table}}
\renewcommand{\theequation}{\thesection\arabic{equation}}

\end{appendices}

\end{document}